# Emergence of excitonic superfluid at topological-insulator surfaces

Yasen Hou[1], Rui Wang[2], Rui Xiao[1], Luke McClintock[1], Henry Clark Travaglini[1], John P. Francia[1], Harry Fetsch[3], Onur Erten[4], Sergey Y. Savrasov[1], Baigeng Wang[5], Antonio Rossi[1,6], Inna Vishik[1], Eli Rotenberg[6] & Dong Yu[1*]

[1]*Department of Physics, University of California, Davis, California 95616, USA*

[2]*Department of Physics and Astronomy, Shanghai Jiao Tong University, Shanghai 200240, China*

[3]*Department of Physics, Harvey Mudd College, Claremont, California 91711, USA*

[4]*Department of Physics, Arizona State University, Arizona 85281, USA*

[5]*Department of Physics, Nanjing University, Jiangsu 210008, China*

[6]*Advanced Light Source, Lawrence Berkeley National Laboratory, Berkeley, California 94720, USA*

*\*e-mail: yu@physics.ucdavis.edu*

**Excitons are spin integer particles that are predicted to condense into a coherent quantum state at sufficiently low temperature, and exciton condensates can be realized at much higher temperature than condensates of atoms because of strong Coulomb binding and small mass. Signatures of exciton condensation have been reported in double quantum wells[1-4], microcavities[5], graphene[6], and transition metal dichalcogenides[7]. Nonetheless, transport of exciton condensates is not yet understood and it is unclear whether an exciton**

condensate is a superfluid[8,9] or an insulating electronic crystal[10,11]. Topological insulators (TIs) with massless particles and unique spin textures[12] have been theoretically predicted[13] as a promising platform for achieving exciton condensation. Here we report experimental evidence of excitonic superfluid phase on the surface of three-dimensional (3D) TIs. We unambiguously confirmed that electrons and holes are paired into charge neutral bound states by the electric field independent photocurrent distributions. And we observed a millimetre-long transport distance of these excitons up to 40 K, which strongly suggests dissipationless propagation. The robust macroscopic quantum states achieved with simple device architecture and broadband photoexcitation at relatively high temperature are expected to find novel applications in quantum computations and spintronics.

The gapless TI surface state is protected against backscattering and has a linear energy dispersion with massless fermions. Though free-fermions have been extensively studied in TIs, much less work is carried out to understand interacting systems[14,15], where electron-electron interaction may lead to emerging quasi-particles. Photoexcited electrons and holes in TIs relax to the surface Dirac cones on sub-picosecond (ps) timescales, while further carrier recombination can be much slower, ranging from a few ps to over 400 ps[16-18]. The long-lived population inversion allows electrons and holes in the transient state to form pairs (Fig. 1a). Because of the small effective mass, excitons in Dirac materials are expected to have long de Broglie wavelength and high transition temperatures $(T_c)$[19]. The figure of merit for exciton formation in materials is $\alpha = \frac{E_C}{E_K}$, where $E_C$ is the Coulomb energy and $E_K$ is the electron kinetic energy. The linear dispersion of the TI surface state results in $\alpha = e^2/\epsilon\hbar v_F$, where $e$ is the electron charge, $\epsilon$ is the dielectric constant of the material, and $v_F$ is the Fermi velocity[20]. The two-dimensional (2D) surface state of a 3D TI, with a single non-degenerate Dirac cone, relatively low $v_F$ (compared to graphene), and reduced $\epsilon$ at surface, has been theoretically identified as a promising candidate for realizing high-$T_c$ exciton condensates[13]. In addition, the topological nature of the band structure may create exotic spin texture to the excitonic quantum state. The spin-momentum locking demands that the ground state of excitons must be a spin-triplet $p$-wave which spontaneously breaks time reversal symmetry[21].

Experimental evidence of exciton condensation in other systems has often been obtained from spatially resolved photoluminescence[2,3,5]. However, TIs are in general not strong photo-emitters because of their gapless surface states. Here we apply scanning photocurrent microscopy (SPCM)[22,23] to visualize the propagation of exciton condensates in 3D TIs. $Bi_{2-x}Sb_xSe_3$ nanoribbons were grown by chemical vapour deposition (CVD)[24], with x = 0.38 determined from energy-dispersive X-ray spectra (EDS). The experimental setup is shown in Fig. 1e inset, where a nanoribbon is electrically connected by two metal contacts (Fig. 1b) and is locally excited by a focused laser. As the laser beam is raster scanned on the device substrate, the photo-induced current is measured as a function of laser position and plotted into a 2D map (Fig. 1c). At room temperature, photocurrent is only observed when the laser is focused close to the contacts, caused by photo-thermoelectric effects. As the temperature is reduced, the photocurrent becomes much stronger and its direction is reversed. And the photocurrent can be observed even when the laser is focused outside the channel and far away from the contacts. Strikingly, below 40 K the photocurrent barely decays even when the photoexcitation position is more than 200 μm away from the contact (Fig. 1c, d). The photocurrent decay length ($L_d$) at various temperature is determined by fitting photocurrent distributions with a hyperbolic function $I(x_0) = A\cosh\left(\frac{x_0-L}{L_d}\right)$, where $x_0$ is the excitation position and $L$ is the channel length (more on $L_d$ extraction and error analysis in Methods). Remarkably, $L_d$ is below 3 μm at 200 K but increases to 0.9 ± 0.3 mm at 7 K, accompanied by an internal quantum efficiency (IQE) as high as 60% (Fig. 1e).

Photocurrent generation with local photoexcitation has been extensively studied[22,23] and is generally understood by the diffusion and separation of photoexcited carriers. In this model, $L_d$ = $\sqrt{D\tau}$, where $D = \mu k_B T/e$ is the diffusion coefficient and $\tau$ is the lifetime of carriers. Transient photocurrent measurements showed $\tau$ = 15 ± 5 ns in our samples (Extended Data Fig. 6), though this value is likely limited by the temporal resolution of our instruments. Taking $\tau$ = 20 ns and $L_d$ = 0.9 mm, we estimate a lower limit of mobility $\mu = eL_d{}^2/\tau k_B T \approx 6 \times 10^4$ m$^2$/(V s) at 7 K. This is 6 orders of magnitude higher than the field-effect mobility determined in our devices ($\mu$ = 0.037 m$^2$/(V s)), and 4 orders of magnitude higher than the highest reported mobility in Bi$_2$Se$_3$ ($\mu$ ~ 1 m$^2$/(V s)[25,26]). Note that though the electron back scattering at the surface of a 3D TI is forbidden, scattering into other angles is possible[27,28], resulting in finite carrier mobility. Therefore, the long $L_d$ and the implied gigantic mobility suggest a highly dissipationless state fundamentally different from free fermions, thus signalling an excitonic superfluid. In this picture, the excitons propagate across the TI surface ballistically and are separated near the contacts. It is known that Bi$_2$Se$_3$ makes Ohmic contact to metals but strong band bending of hundreds of meV exists at the junction[29] and facilitates efficient charge transfer (Extended Data Fig. 3). The high IQE value at low temperature indicates that a large fraction of photoexcited carriers condense in the superfluid state. Both $L_d$ and IQE increase rapidly as temperature decreases and saturate below 40 K (Fig. 1e), which corresponds to a $T_c$ value significantly higher than most of the previous reports[1-6].

To confirm the excitonic nature of photogenerated carriers, we measured $L_d$ as a function of electric field applied along the nanoribbons. Free fermions are charged particles that drift under electric field, so $L_d$ is expected to increase along the electric force. At high electric field, $L_d$ is proportional to the field ($L_d = \mu\tau E$)[30], as experimentally demonstrated in semiconductor nanowires[31,32] and halide perovskites[33]. On the contrary, excitons are charge neutral, so $L_d$ is not expected to change under external electric field. Our measurements clearly showed that $L_d$ remains constant and independent of electric field (Fig. 2). Note that the applied electric field here is much lower than that needed to separate excitons (Methods). These measurements clearly confirm that the photogenerated carriers form excitons.

Long $L_d$ is only observed in Sb-doped $Bi_2Se_3$ samples, in which the Fermi level ($E_F$) is close to the Dirac point evidenced by the ambipolar gate dependence (Fig. 3d) and micro-angle-resolved photoemission spectroscopy (ARPES)[34]. Micro-ARPES spectra of these nanostructures have demonstrated clear Dirac cones and indicated that the samples are slightly $n$-doped relative to the Dirac point, but with $E_F$ below the bulk conduction band (Extended Data Fig. 2). $L_d$ in samples with low Sb doping is shorter than that with more Sb (Extended Data Fig. 7). In pure $Bi_2Se_3$ that is degenerately $n$-doped, photocurrent with much lower magnitude is observed solely near the contacts (Fig. 3a). This explains why nonlocal photocurrent has not been reported in TIs, though photocurrent mapping in TIs has been studied previously[35-37]. The shorter $L_d$ in $n$-doped TIs is likely caused by faster exciton recombination and/or stronger electric field screening that greatly suppress the exciton formation.

The photocurrent distributions are also measured as a function of light polarization, excitation wavelength, gate voltage, and light intensity (Fig. 4). Both circularly and linearly polarized laser beams are applied but the resulting photocurrent distributions are independent of the polarization because of normal incidence of the laser[37]. $L_d$ is also independent of excitation wavelength (Fig. 4a, d), which rules out the possibility of surface plasmon polariton (SPP) since the SPP propagation length is expected to be wavelength dependent[38]. In addition, it confirms that the second Dirac cone 1.5 eV above the conduction band edge is not involved in the exciton formation[39]. $L_d$ and IQE can be greatly modulated by gate voltage ($V_g$). As $V_g$ lowers $E_F$, photocurrent first increases slightly, presumably due to weaker screening. As $V_g$ is further increased, it tunes the TI from $n$-type to $p$-type. Both photocurrent and $L_d$ drop sharply (Fig. 4b, e), which we attribute to faster recombination caused by the mixing of surface states and bulk valence band in $p$-type TIs, as the Dirac point is close to the bulk valence band in $Bi_2Se_3$. Furthermore, the photocurrent decays faster at higher laser intensity (Fig. 4c, f), indicating that stronger screening at high intensity makes exciton formation more difficult. The theoretically estimated $T_c$ as a function of excitation power (Methods) is in good agreement with the experimental observation (Extended Data Fig. 8, 9).

Different theories suggest that exciton condensates can be either an insulating[11] or a superfluid state[19,40]. The observed highly dissipationless exciton transport in TIs provides strong evidence supporting superfluidity. Furthermore, Fig. 1e shows that $L_d$ decreases gradually when temperature is increased above $T_c$, indicating that phase transition is not sharp. This can be understood by considering Kosterliz-Thouless (KT) phase transition[41], which predicts the phase transition of a 2D system into a superfluid under the KT transition temperature $T_{KT}$, despite the absence of conventional long-range order in 2D. In this theory, the vortex excitations emerge and become closely bound for $T < T_{KT}$, resulting in a coherent state that displays frictionless exciton transport[41,42]. For $T > T_{KT}$, the vortices are unbound and the correlation length scales exponentially with $T$[43] ($\xi = c_1 \exp[\frac{c_2}{T-T_{KT}}]^{1/2}$). The characteristic length of exciton transport, i.e., $L_d$, is then expected to have a similar scaling and decrease gradually at higher temperature. The $T_{KT}$ values estimated by the Hartree-Fock mean-field calculations are in good agreement with the experimental observation (Extended Data Fig. 9). Finally, the observed topological exciton condensates are unique in that they are from direct excitons at TI surface, while excitons in previous systems are indirect with electrons and holes either spatially separated by an insulating layer[1-4,6] or at energy minima with different momenta[7]. Consequently, the coherent macroscopic quantum states are robust and can be realized in simple devices that do not involve complex structures, enabling widespread applications in quantum computations and spintronics.

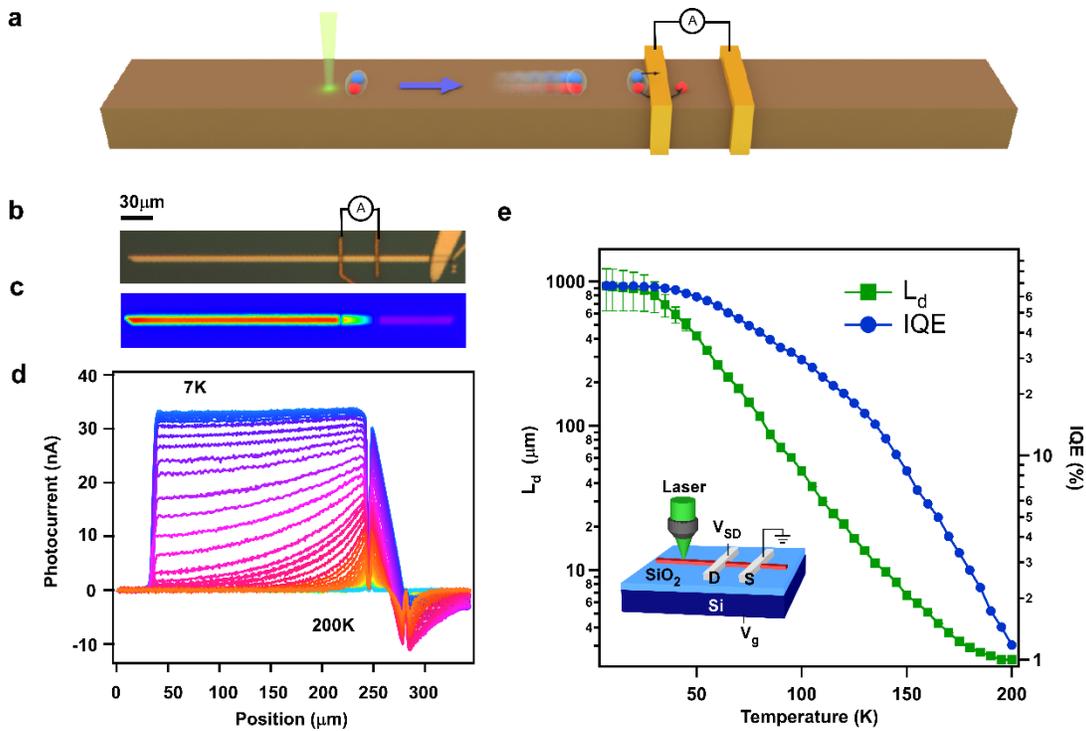

**Figure 1: Nonlocal photocurrent generation in a TI nanoribbon. a,** Schematic of exciton transport in TIs. Electrons and holes, denoted by blue and red balls respectively, are bound and travel ballistically at TI surface until being separated at the metal contact. **b,** Optical image of a Sb-doped $Bi_2Se_3$ nanoribbon ($305 \times 6.5 \times 0.13 \ \mu m^3$) in contact with Cr/Au electrodes. The far right end of the nanoribbon is in contact with another TI nanoplate. **c,** A photocurrent map collected by scanning a focused laser at normal incidence at 7 K and zero source-drain and gate biases. Laser power is 166 nW. **d,** Photocurrent distributions along the nanoribbon axis at various temperatures. **e,** $L_d$ and IQE as a function of temperature. IQE (electron collected per absorbed photon) is calculated from the photocurrent and laser power considering a reflectance of 30%. The uncertainty of $L_d$ becomes large at low temperature because $L_d$ is several times larger than $L$. Inset: SPCM setup.

**a** **b**

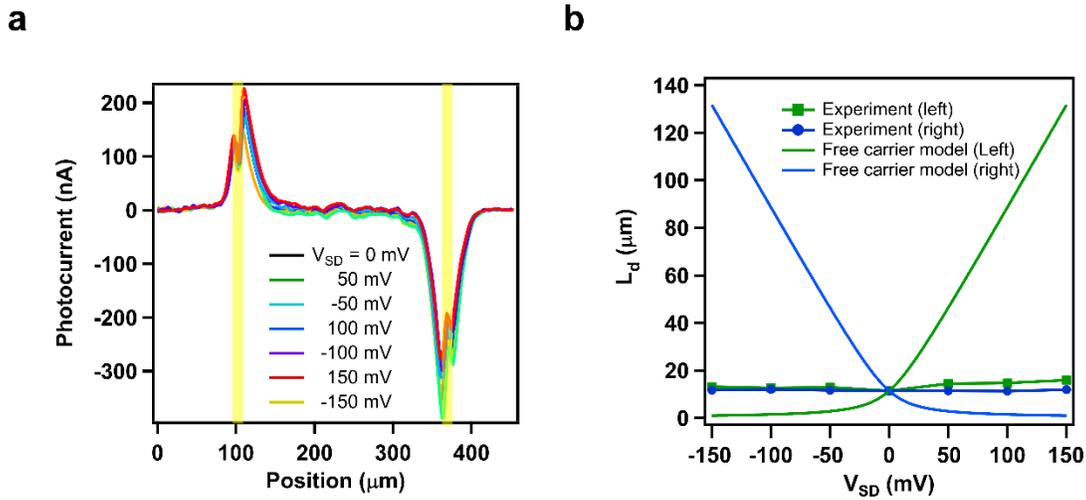

**Figure 2: Electric field independent photocurrent profiles at 7 K.** The dark current induced by source-drain bias ($V_{SD}$) is subtracted from the total current. $V_g$ = -75 V is applied to shorten $L_d$ in order to observe possible electric field induced changes. **a,** Photocurrent as a function of laser excitation position along the nanoribbon axis at various $V_{SD}$. Laser power is 20 μW. Vertical yellow lines indicate the contacts. **b,** $L_d$ as a function of $V_{SD}$. Solid lines represent the photocurrent decay lengths predicted by the free carrier model using $L = \dfrac{2L_D{}^2}{\sqrt{L_E{}^2 + 4L_D{}^2} \mp L_E}$ and $L_E = eEL_D{}^2/k_BT$ as in reference 30. $L_D$ is determined to be 11.4 μm at $V_{SD}$ = 0 V. The field independent $L_d$ indicates carriers are bound into charge neutral excitons.

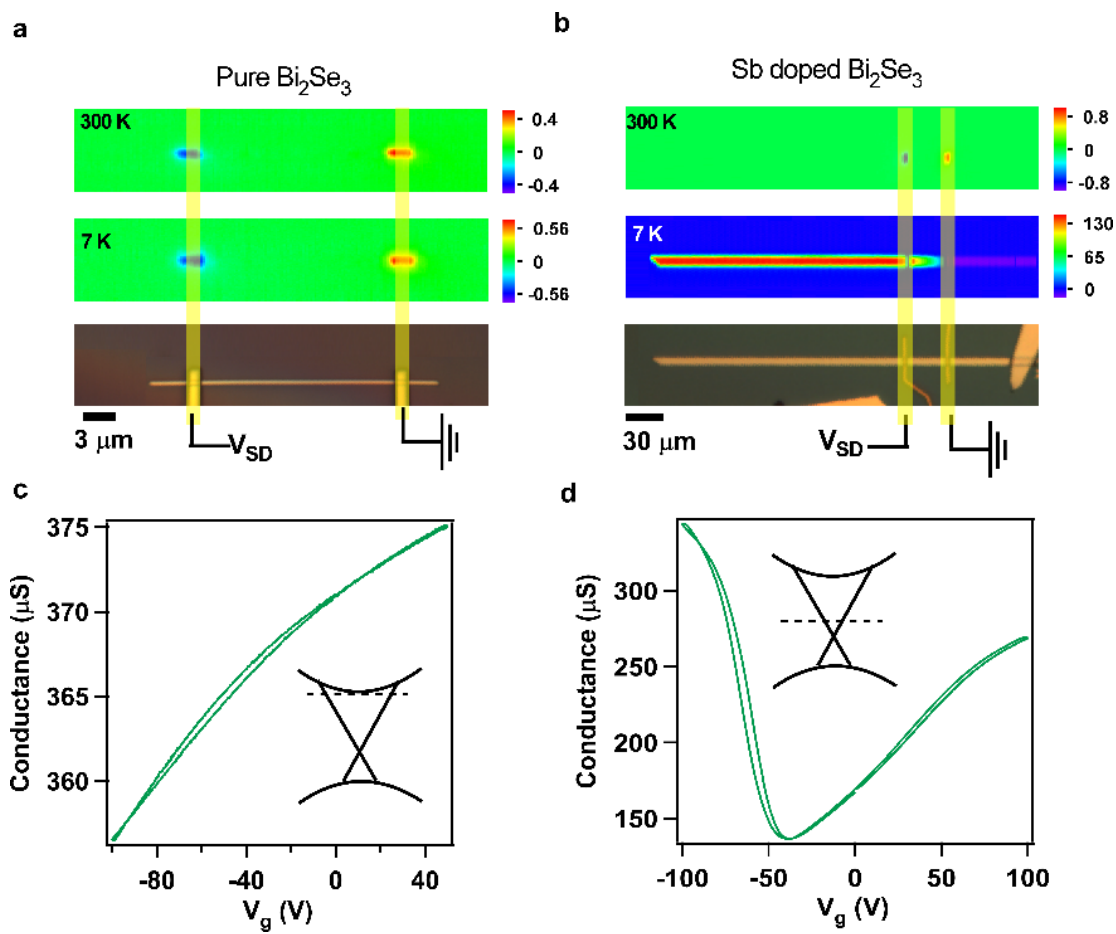

**Figure 3: Doping dependent photocurrent profiles. a-b,** Photocurrent and optical images, where vertical yellow lines indicate the contacts. Colour scales are current in nA. Laser power is 723 nW. **c-d,** Gate dependent conductance measured in the dark at 7 K. Insets: band diagrams showing $E_F$ positions. **a** and **c** are for pure $Bi_2Se_3$, where $E_F$ is close to the conduction band. Field effect mobility and electron concentration are estimated to be $\mu$ = 329 $cm^2$/Vs, n = 3.25×$10^{18}$ $cm^{-3}$. The photocurrent is only observed when excitation is close to the contacts. **b, d,** Sb doping lowers $E_F$ as evidenced by ambipolar conduction. $\mu$ = 371 $cm^2$/Vs, n = 9.3×$10^{16}$ $cm^{-3}$ for electrons.

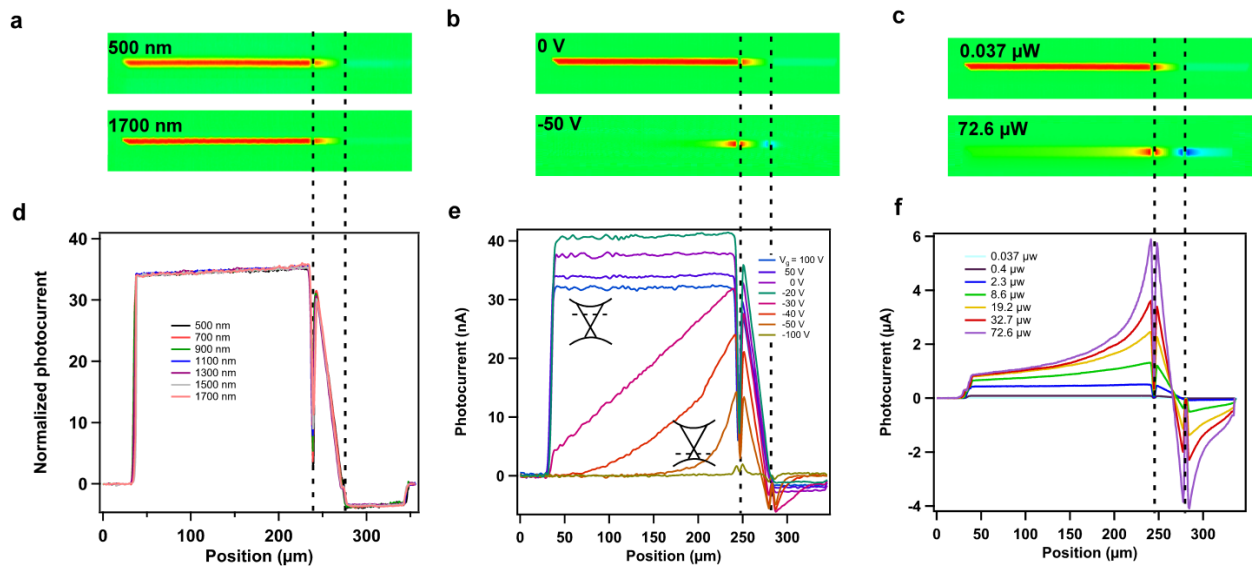

**Figure 4: Effects of wavelength, gate voltage, and laser power on photocurrent decay lengths in Sb-doped Bi$_2$Se$_3$.** The dashed lines indicate the contacts. The measurements are carried out at 7 K and zero source-drain and gate biases. **a-c,** 2D photocurrent maps. **d-f,** line cuts along the nanoribbons axis. **a, d,** Wavelength dependence. Laser power from 77 nW to 280 nW was used for different wavelengths to maintain the same exciton injection rate. **b, e,** Gate voltage dependence. Inset, band diagrams showing $E_F$ position. Laser power is 166 nW. **c, f,** Laser power dependence.

**Acknowledgements** This work was supported by National Science Foundation Grant DMR-1838532 and DMR-1710737. S.Y.S was supported by National Science Foundation Grant DMR-1411336. This research used the Molecular Foundry and the Advanced Light Source, which are US Department of Energy Office of Science User Facilities under contract no. DE-AC02-05CH11231. H.F. acknowledges the U.S. National Science Foundation Research Experiences for Undergraduates (REU) program under Grant No. PHY-1560482.

**Author contribution** D. Y. and Y. H. designed the experiments. Y. H. synthesized $Bi_2Se_3$ nanoribbons and performed the measurements. R. X., L. M., H. C. T., J. F., and H. F. assisted the synthesis and measurements. A. R., I. V., and E. R. performed micro-ARPES measurements. R. W., O.E., and S. S. performed theoretical calculation. All the authors analysed the data. D. Y., Y. H., R. W., O. E., and I. V. co-wrote the paper.

**Methods**

**Nanoribbon Growth and Device Fabrication.** The CVD growth was carried out in a Lindberg Blue M tube furnace, following similar procedures as in previous work[24]. The system was first evacuated to a base pressure of 30 mTorr and Ar was then injected and a room pressure was maintained. For a typical growth, 116 mg of $Bi_2Se_3$ powder (99.999%, Alfa Aesar) was mixed with 20-35 mg of Sb powder (99.999%, Alfa Aesar) and placed in a small quartz tube at the centre of the tube furnace. 250 mg of Se pellets (99.999%, Johnson Matthey Inc.) were placed in another quartz tube upstream by a distance of 16 cm. A silicon substrate was placed 14 cm downstream from the centre of the furnace. The surface of the silicon substrate was coated with 10 nm of Au as a catalyst by electron beam evaporation. The temperature at the centre of the furnace was 680 °C, the Ar flow rate was 150 sccm (standard cubic centimetres per minute) and the growth time was 5 h. After that, the furnace was cooled down to room temperature over approximately 3 hours. The growth yields both nanoribbons and nanoplates (Extended Data Fig. 1a). The as-grown nanoribbons were then transferred to 300 nm $SiO_2$ covered Si substrates, where single nanoribbon field effect transistor (FET) devices were fabricated using a standard electron beam lithography process. Top metal contacts (10 nm Cr / 290 nm Au or 10 nm Ti / 290 nm Au) were made using an electron beam evaporator (CHA) or a sputterer (Lesker). A typical device is shown in Extended Data Fig. 1b,c.

**Optoelectronic measurements.** The low temperature measurements were performed in a cryostat (Janis ST-500). Current-voltage curves were measured through a current preamplifier (DL Instruments, model 1211) and a NI data acquisition system. Scanning photocurrent microscopy (SPCM) measurements were performed using a home-built setup based upon an Olympus microscope. Briefly, a 532 nm CW laser or a tuneable laser (NKT SuperK plus) was focused by a 10× N.A. 0.3 objective lens to a diffraction limited spot with a size of ~3 μm and raster scanned on a planar nanoribbon device by a pair of mirrors mounted on galvanometers, while both reflectance and photocurrent were simultaneously recorded to produce a 2D maps. The laser power was controlled by a set of neutral density (ND) filters and was measured by a power meter underneath the objective lens. Fast photoresponse measurements were performed using a pulsed laser (Thorlabs 450 nm, pulse width 10-40 ns), high-speed amplifiers (Femto DHPCA-100), and a digital oscilloscope. The results in Fig. 1, 3, 4 (except Fig. 3a, c) are obtained from one device for consistency, while the general trends are highly repeatable in more than 10 devices measured to date.

**Micro-ARPES.** Micro-ARPES experiments were performed at the Microscopic and Electronic Structure Observatory (MAESTRO) beamline at the Advanced Light Source. Samples were removed from growth chamber, sealed in an argon gas environment, and inserted into the micro-ARPES UHV end-station with a base pressure better than $5 \times 10^{-11}$ mbar via attached glovebox. ARPES data were collected at 70 K using a hemispherical Scienta R4000 electron analyzer, 100 eV photon energy, and a beam size of 5 ×10 μm². A typical micro-ARPES spectrum of a Sb-doped $Bi_2Se_3$ nanoplate is shown in Extended Data Fig. 2.

**Extraction of photocurrent decay length.** A hyperbolic function must be used instead of exponential when $L$ is comparable or shorter than $L_d$. We have fitted the photocurrent distributions by an exponential function $I(x_0) = A\exp(-\frac{x_0}{L_d})$ or a hyperbolic function $I(x_0) = A\cosh\left(\frac{x_0-L}{L_d}\right)$, where $L$ is the channel length and $L_D$ is the photocurrent decay length. The two fittings are similar when $L_d \ll L$, but the hyperbolic function fits more accurately when $L_d$ is comparable or longer than $L$ (Extended Data Fig. 4). We justify the hyperbolic fitting below. The steady state continuity equation describing exciton concentration is,

$$D\frac{\partial^2 n}{\partial x^2} - \frac{n}{\tau} + G\delta(x-x_0) = 0 \qquad (1)$$

where $D$ is the diffusion coefficient, $\tau$ is the lifetime, $G$ is the generation rate proportional to laser power, and the device geometry is shown in Extended Data Fig. 5. The local laser generation is considered as a delta function. For boundary conditions, we assume all excitons separated at contact, so $n$ drops to zero at $x = 0$. The excitons cannot flow out of the nanoribbon, so $D\frac{\partial n}{\partial x}$ drops to zero at $x = L$. The exciton concentration is continuous at $x = x_0$ and its derivative follows a relation that can be found out by integrating equation (1). The solution of exciton concentration then can be found out, from which we can calculate current distribution as,

$$I(x_0) = eD\frac{\partial n}{\partial x}(x = 0) = eG\frac{\cosh(\frac{x_0-L}{L_d})}{\cosh(\frac{L}{L_d})} \qquad (2)$$

where $L_d = \sqrt{D\tau}$. The above derivation considers a diffusive process. For a ballistic process, if we assume that the injected excitons have equal probability of moving left or right and bounce back at the end of the nanoribbon without loss, we can derive a similar photocurrent distribution,

$$I(x_0) = I_L + I_R = \frac{eG}{2}\left(e^{-\frac{x_0}{L_d}} + e^{-\frac{2L-x_0}{L_d}}\right) = eGe^{-\frac{L}{L_d}}\cosh(\frac{x_0-L}{L_d}) \qquad (3)$$

where $I_L$ is the current contributed by the excitons moving left from the injection point to contact and $I_R$ is the current contributed by the excitons moving right from the injection point and then bounced back at the end of the nanoribbon. $L_d = v\tau$ is the average distance that an exciton condensate can travel ballistically before recombination. The exciton velocity $v$ is expected to be close to the Fermi velocity of the massless electrons at the TI surface. Therefore, we showed the current follows a hyperbolic function with the excitation position, in both diffusive and ballistic cases.

Error analysis. The main error in $L_d$ originates from error in photocurrent and laser power. The photocurrent error is about 1%. Because of the large scan areas, the incident power is not uniform and the photocurrent presented has already been corrected with the power at different injection position. The uncertainty in power measurement then also affects the corrected photocurrent and is estimated to be 0.5%. From equation (2) or (3), it is easy to find $L_d = \frac{L}{\cosh^{-1}(z)}$, where $z = I(x_0 = 0)/I(x_0 = L)$. We then propagate these error sources to estimate the error in $L_d$,

$$\Delta L_d = \left| \frac{d}{dz} \left[ \frac{1}{\cosh^{-1}(z)} \right] \right| L \Delta z = \frac{1}{[\cosh^{-1}(z)]^2 \sqrt{z^2 - 1}} L \Delta z \tag{4}$$

where $\frac{\Delta z}{z} = \sqrt{2 \left[ \left( \frac{\Delta J}{J} \right)^2 + \left( \frac{\Delta P}{P} \right)^2 \right]} \approx 1.6\%$. Both currents are corrected by power, i.e., $I = J/P$ where J is the uncorrected current and P is laser power. The factor of 2 is because $z$ is the ratio of two currents at $x_0 = 0$ and $x_0 = L$. The error is large when $L_d$ is long at low temperature since $z$ is close to 1. The error of $L_d$ at different temperature is shown in Fig. 1e.

**Theoretical calculation of critical temperature.** We assume a quasi-equilibrium state where the excited electrons and holes stay in the upper and lower Dirac cones respectively with well-defined chemical potentials $\mu_+$ and $\mu_-$. The noninteracting Hamiltonian can be written as,

$$H_0 = \sum_{\mathbf{k}} (v_F k - \mu_+) c_{\mathbf{k},+}^\dagger c_{\mathbf{k},+} + \sum_{\mathbf{k}} (-v_F k - \mu_-) c_{\mathbf{k},-}^\dagger c_{\mathbf{k},-} \tag{5}$$

where $c_{\mathbf{k},+}$ and $c_{\mathbf{k},-}$ are annihilation operators for electrons in the upper and lower Dirac cones respectively. In the picture of electrons, we consider a Coulomb interaction,

$$H_{int} = \sum_{\mathbf{k},\mathbf{k}',\mathbf{q}} \frac{2\pi e^2}{(\kappa + q)\epsilon} c_{\mathbf{k}+\mathbf{q},\sigma}^\dagger c_{\mathbf{k}'-\mathbf{q},\sigma'}^\dagger c_{\mathbf{k}',\sigma'} c_{\mathbf{k},\sigma} \tag{6}$$

where the screening effect has been taken into account with the screening wave number $\kappa = \alpha(k_F^+ + k_F^-)$, where $\alpha = e^2/\epsilon \hbar v_F$ and $k_F^\pm$ is the wave number at the Fermi surfaces of the electron and hole pockets. The Coulomb repulsive interaction between electrons, equivalent to the attractive interaction between electrons and holes, can generate excitons and gap out the Dirac state. Similar to reference [13], we perform a mean-field calculation of the possible Cooper instability, analogous to the BCS theory but in the particle-hole channel. Specifically, we introduce mean-field order parameter in the particle-hole channel with $\triangle_{\mathbf{k}} = \sum_{\mathbf{k}'} V_{\mathbf{k}-\mathbf{k}'} < c_{\mathbf{k}',-}^\dagger c_{\mathbf{k}',+} >$, where $V_{\mathbf{k}-\mathbf{k}'} = \frac{2\pi e^2}{(\kappa + |\mathbf{k}-\mathbf{k}'|)\epsilon} = \frac{2\pi \alpha \hbar v_F}{\kappa + |\mathbf{k}-\mathbf{k}'|}$. The self-consistent equation can be obtained by minimizing the energy of the mean-field ground state as,

$$\triangle_{\mathbf{k}} = \frac{1}{2} \int \frac{d\mathbf{k}'_{\epsilon\Omega}}{(2\pi)^2} V_{\mathbf{k}-\mathbf{k}'} \triangle_{\mathbf{k}'} \frac{n_F(\xi_{\mathbf{k}'} - \mu_1) - n_F(-\xi_{\mathbf{k}'} - \mu_1)}{\xi_{\mathbf{k}'}} \tag{7}$$

where $\xi_\mathbf{k} = \sqrt{(\hbar v_F k - \mu_2)^2 + |\triangle_\mathbf{k}|^2}$, $\mu_{1,2} = (\mu_+ \pm \mu_-)/2$, and the integral is restricted to a momentum cutoff set by $\Omega$. Since the photons generate equal number of electrons and holes and the original chemical potential of the sample in the dark is close to the Dirac point, we expect that the electron and hole Fermi surfaces enjoy a perfect nesting with chemical potential $\mu_\pm = \pm\mu$ (therefore $\mu_1 = 0$, $\mu_2 = \mu$). Indeed, the self-consistent solution of the equation with a $\mathbf{k}$-resolved order parameter indicates that the ground state can develop a full excitonic gap due to BCS condensation.

Furthermore, if one considers a spatially-resolved complex order parameter, the thermal fluctuation is found to generate vortices at finite temperature that are more stable than the real order parameter solutions. Meanwhile, the KT transition takes place characterizing the phase transition. The KT temperature is the energy scale for the proliferation of vortices and antivortices. To calculate the KT temperature, we first evaluate the exciton current $\mathbf{j_Q}$ with a small momentum $\mathbf{Q}$ of the excitons. The superfluid density $\rho_s$ can then be obtained from $\mathbf{j_Q} = e\,\rho_s\mathbf{Q}/\hbar$. The superfluid density directly determines the transition temperature through $T_{KT} = \pi\rho_s(T_{KT})/2$. Since $\rho_s$ is dependent on $\triangle_\mathbf{k}$ which again relies on $\mu$, $T_{KT}$ is a function of $\mu$, and therefore dependent on the laser intensity. When lowering the chemical potential $\mu$, the obtained $T_{KT}$ first increases because of the weaker screening and then decreases due to the lower density of states at small $\mu$. Since $\mu$ is proportional to the laser intensity, these results reveal the trend of the growth of $T_{KT}$ with lowering intensity, and are qualitatively in agreement with the experimental observation (Extended Data Fig. 8). However, the evolution of $T_{KT}$ at very low laser intensity is not experimentally observable due to low signal. The obtained mean-field phase diagram as well as the KT temperature are shown in Extended Data Fig. 9, where an excitonic condensate supporting the long-range photocurrent is stabilized from the electron-hole gas after crossing KT transition curve ($T_{KT}$ as a function of $\mu$). For $\alpha = 0.4$, the maximum $T_{KT}$ is calculated to be 37 K, comparable to the experimental observation. This $\alpha = e^2/\epsilon\hbar v_F$ value is expected to be in the range of 0.1 to 0.5, as $\epsilon$ at low frequency at the TI surface is about one half of the bulk value and is expected to vary from 50 to 10 in the previous reports[25,44].

Finally, we estimate the electric field required to split the excitons. We first calculate the condensation energy $E_{cond}$ of the excitonic phase, i.e., the energy reduction of the excitonic order compared to the original Dirac surface state. $E_{cond} = -3.0$ meV for the optimal $\mu_2$ (0.1 eV) with $\alpha = 0.4$. The Bohr radius can then be estimated by $a_B = \hbar v_F/E_{cond}$, which gives $a_B$ around 110 nm. This leads to an electric field of $E_{cond}/ea_B \sim 2.7 \times 10^4$ V/m, much higher than the maximum electric field (600 V/m) used in Fig. 2.

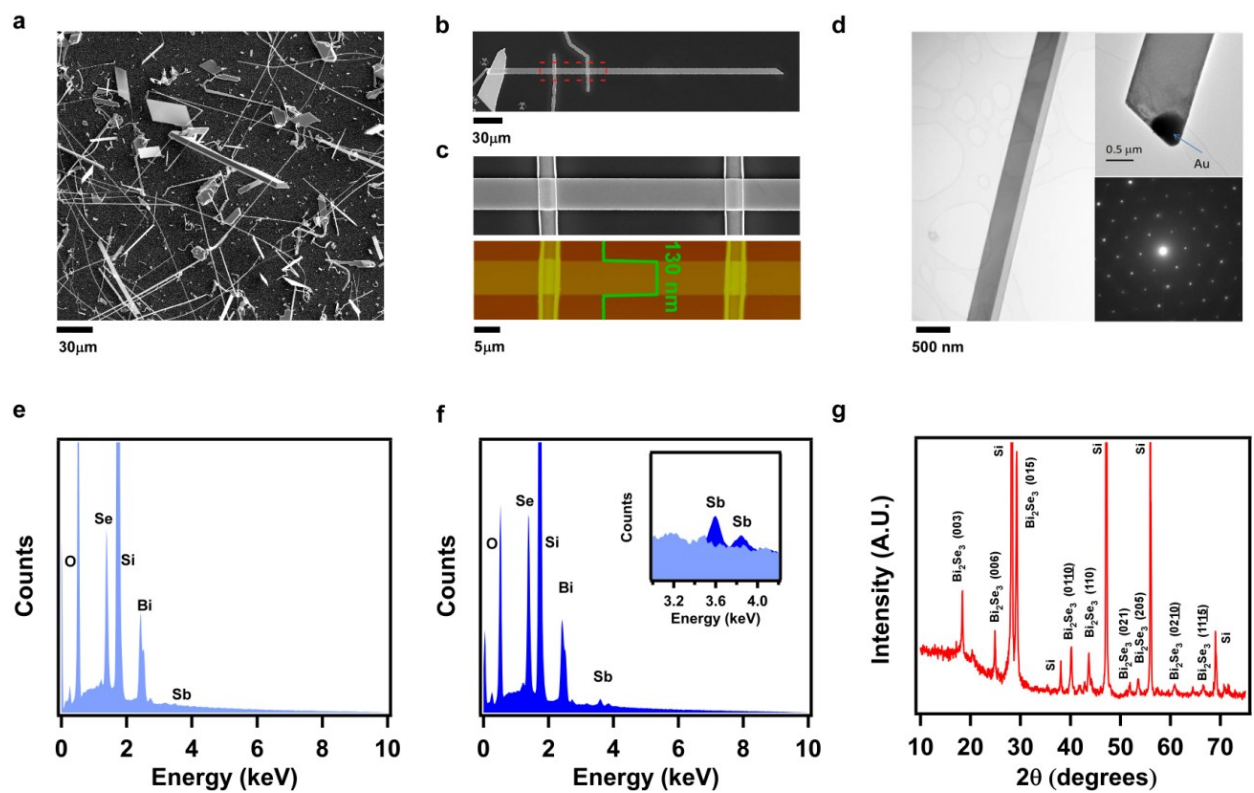

**Extended Data Figure 1: a,** Scanning electron microscopic (SEM) image of as-grown Sb-doped Bi$_2$Se$_3$ nanoribbons and nanoplates. **b,** SEM image of the nanoribbon device in Fig. 1. **c,** Zoom-in SEM and atomic force microscopic (AFM) images of the part in the red square of **b**. **d,** Transmission electron microscopic (TEM) images of Sb-doped Bi$_2$Se$_3$ nanoribbons. Upper inset shows the presence of Au at the tip of a nanoribbon, demonstrating the vapour-liquid-solid (VLS) growth mechanism. Lower inset shows the selected area electron diffraction patterns. **e-f,** Energy dispersive X-ray spectra (EDS) of pure and Sb-doped Bi$_2$Se$_3$ nanoribbons in devices respectively. **f** inset: zoom-in spectra showing the presence of a Sb peak in Sb doped Bi$_2$Se$_3$. The Sb peak corresponds to an atomic percentage of 7.6% or Bi$_{2-x}$Sb$_x$Se$_3$ with x = 0.38. **g,** X-ray diffraction (XRD) patterns of Sb doped Bi$_2$Se$_3$. Note that the Si peaks are from the substrate.

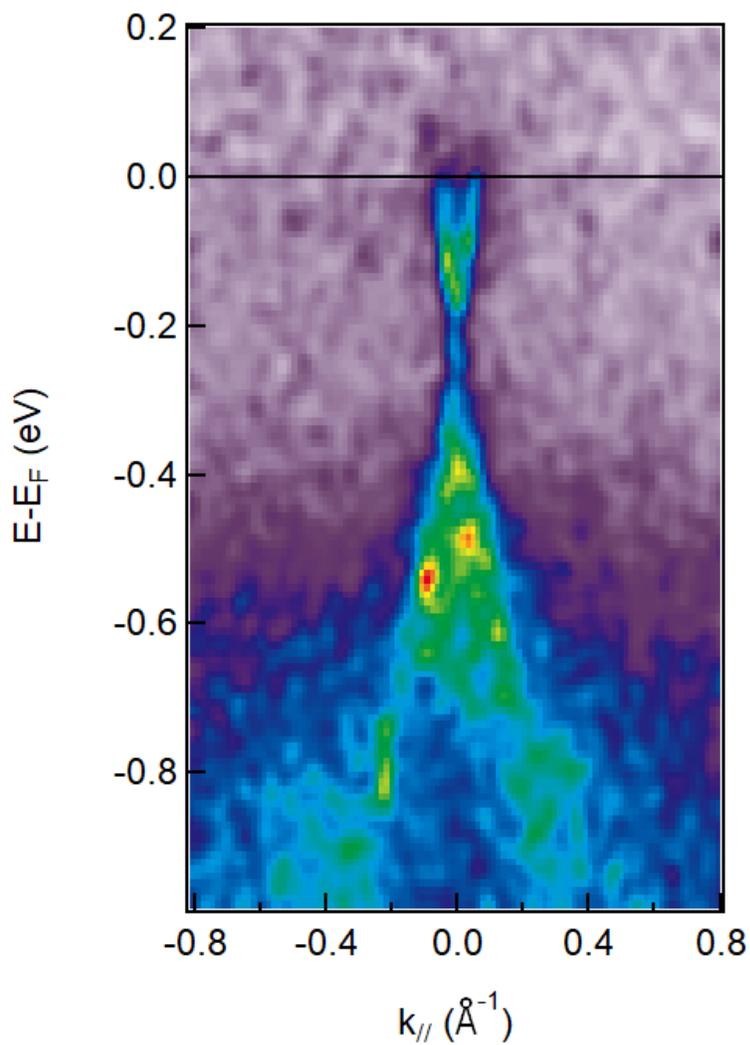

**Extended Data Figure 2:** Micro-ARPES spectrum of a Sb-doped Bi$_2$Se$_3$ nanoplate. Data taken using 5-μm beam spot size on portion of the nanoplate.

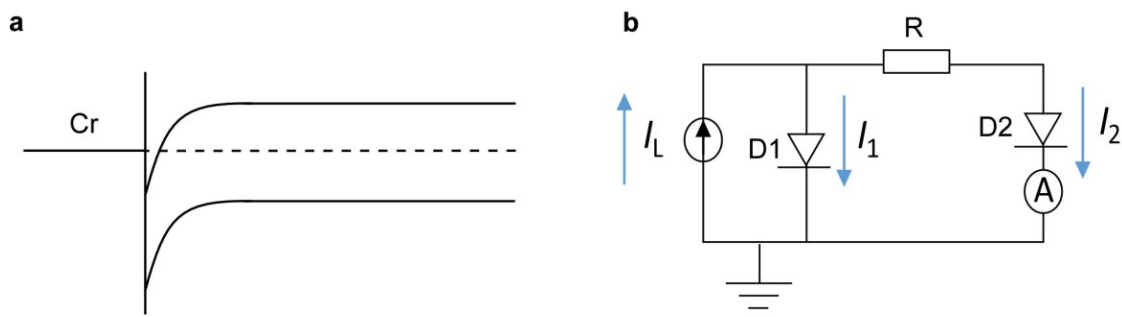

**Extended Data Figure 3: a,** Band diagram and **b,** circuit diagram for understanding current generation in TI nanoribbon devices. The direction of the photocurrent at low temperature is consistent with the downwards band bending towards the metal contact. The charge separation creates a constant current source ($I_L$) as in a photovoltaic cell, leading to the measured photocurrent ($I_2$). The two diodes correspond to the metal-TI junctions at the drain and source contacts, respectively. $R$ is the TI nanoribbon resistance. The current is measured by a preamp as $I_2$. The IQE is $I_2\,h\nu\,/\,eP$, where $P$ is the absorbed power and $h\nu$ is the photon energy. There are three major loss mechanisms: (1) recombination of exciton condensates, (2) recombination of normal excitons, and (3) loss at the junction. (1) is much less than (2) because of the ballistic transport of the condensate. The last mechanism is expected to be small because of the efficient charge transfer at TI and metal junction. Therefore, IQE is an evaluation of the fraction of condensed excitons out of the total excitons.

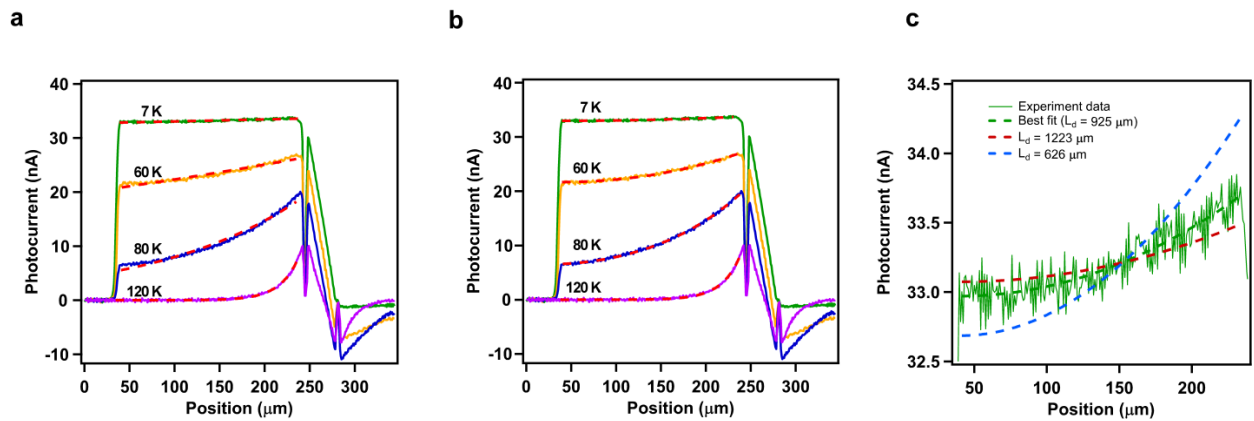

**Extended Data Figure 4: Comparison between exponential and hyperbolic fittings. a,** Photocurrent distribution is fitted by $I(x_0) = A\exp(-\frac{x_0}{L_d})$. **b,** Photocurrent distribution is fitted by $I(x_0) = A\cosh\left(\frac{x_0 - L}{L_d}\right)$. The red dashed curves are fittings. **c,** Zoom-in plot of photocurrent distribution and fitting at 7 K. The data points are comfortably within the fittings with upper and lower limit of $L_d$ calculated from error propagation.

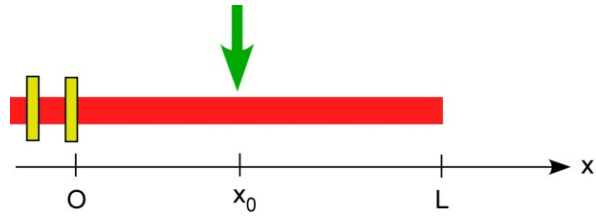

**Extended Data Figure 5:** Configuration of SPCM experimental setup with laser (green arrow) injected at x = $x_0$. The yellow bars indicate the contacts and the red bar indicates the TI nanoribbon. One contact is at x = 0 and the nanoribbon ends at x = L.

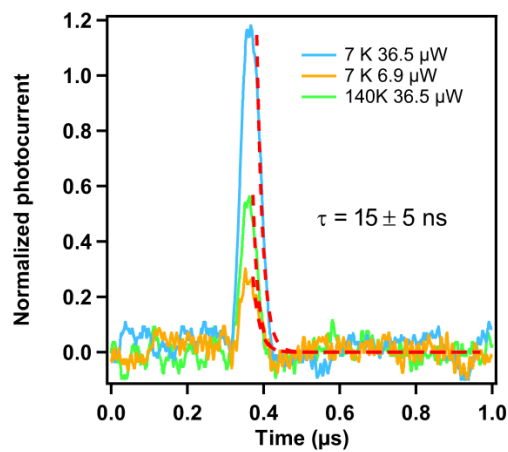

**Extended Data Figure 6: Transient photocurrent measurements**. Photocurrent as a function of time when the device is excited by a pulsed laser of width 40 ns at 7 K and 140 K respectively. The red dashed lines are exponential fittings which yield a decay time of 15 ns. Note this decay time is the upper limit of the real carrier recombination lifetime, as limited by the temporal resolution of the amplifier. The peak powers of the laser are specified in the legend.

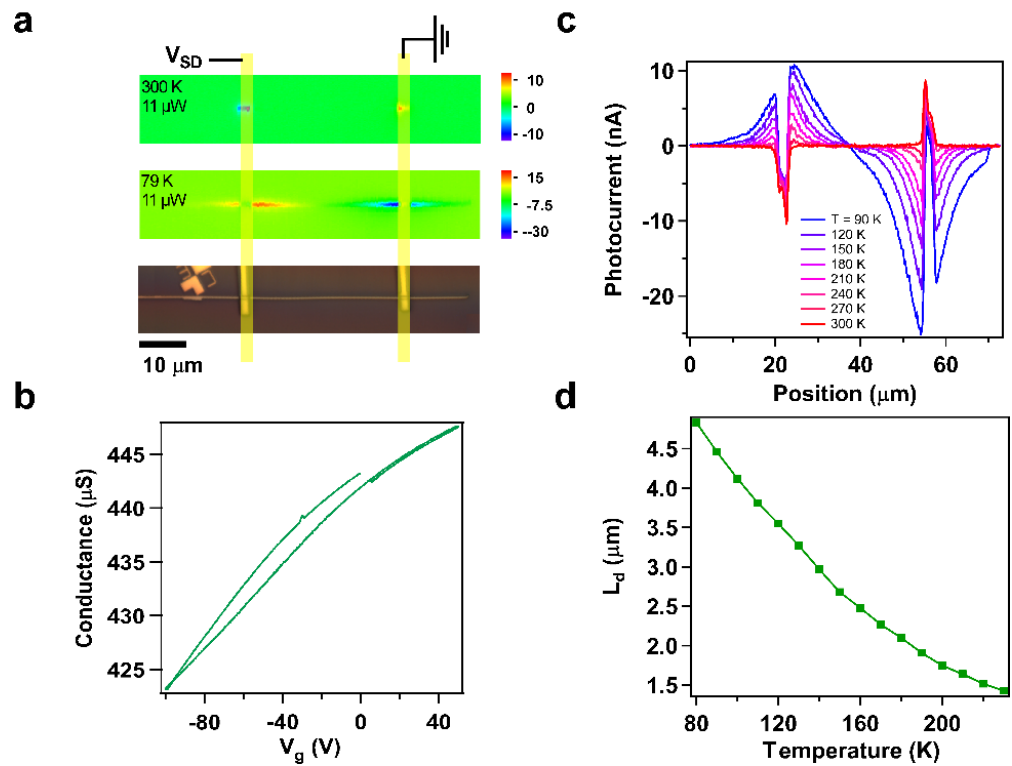

**Extended Data Figure 7**: **Photocurrent mapping for a Bi$_2$Se$_3$ nanoribbon device with low Sb doping.** Sb doping concentration is less than 2.5% according to energy dispersive X-ray spectroscopy (EDS) measurements. **a,** Photocurrent and optical images, where vertical yellow lines indicate the contacts. Colour scales are current in nA. **b,** Gate dependent conductance measured in the dark at 79 K. Field effect mobility and electron concentration are estimated to be $\mu = 1.24 \times 10^3$ cm$^2$/Vs, $n = 1.6 \times 10^{17}$ cm$^{-3}$. **c,** Photocurrent distributions along the nanoribbon axis at various temperatures. **d,** $L_d$ vs. temperature. $L_d$ values are between those of more heavily Sb-doped and pure Bi$_2$Se$_3$ samples.

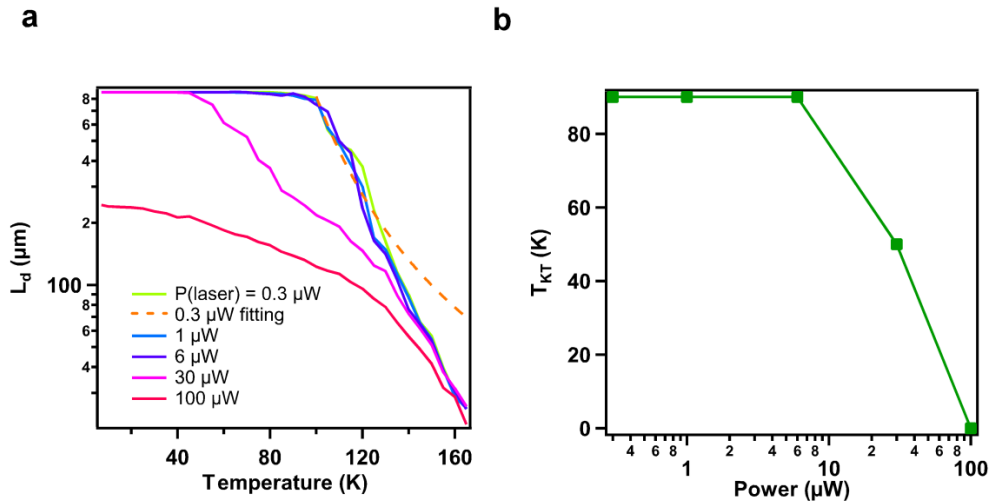

**Extended Data Figure 8: Power dependent $L_d$ and extraction of $T_{KT}$. a,** $L_d$ as a function of temperature at various laser power. The dashed line is the fitting of theoretical expectation $L_d = c_1 e^{[c_2/(T-T_{KT})]^{1/2}}$. The experimental data is in good agreement with the theory near $T_{KT}$ but deviates at higher temperature. **b,** $T_c$ as a function of laser power. $T_{KT}$ is extracted from the turning point in **a**. At higher power, the $T_{KT}$ extracted from the $L_d$ vs T curves decreases rapidly. The red curve at 100 μW does not reach condensate in the experimental temperature range. Here $L_d$ is measured as a function of temperature at different excitation power in a device different from the one shown in Fig. 4c and f. Note that because of the limited length of the nanoribbon, only the lower limit of the $L_d$ can be accurately determined, which we use to represent the $L_d$ value after saturation. The $T_{KT}$ value in this device appears to be higher than that in Fig. 1e.

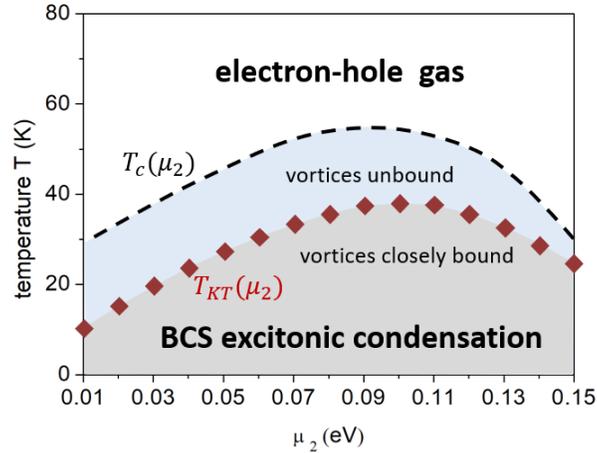

**Extended Data Figure 9: Phase diagram from mean-field calculations.** The electron-hole gas state excited by photons is stable at high temperature, while the system starts to enter into the BCS excitonic condensation phase for temperature lower than the mean-field transition temperature $T_c(\mu_2)$. The calculated KT transition temperature is denoted by the brown data curve $T_{KT}(\mu_2)$. For temperature lower than $T_{KT}(\mu_2)$, the vortices become closely bound and the system displays long-range transport of excitons. The parameters used are $\alpha = 0.4$, $v_F = 5 \times 10^5$ m/s, the energy cutoff of the surface state 0.3 eV, and we assume a perfect nesting with $\mu_1 = 0$.